\documentclass[preprint,showpacs,aps]{revtex4}
\begin{document}
\def\be{\begin{equation}}
\def\ee{\end{equation}}
\def\bea{\begin{eqnarray}}
\def\eea{\end{eqnarray}}
\title{Geometrothermodynamics of black holes}
\author{Hernando Quevedo}
\email{quevedo@nucleares.unam.mx}
\affiliation{ Instituto de Ciencias Nucleares\\
Universidad Nacional Aut\'onoma de M\'exico  \\
A.P. 70-543   \\
04510 M\'exico D.F., MEXICO}

\begin{abstract}

The thermodynamics of black holes is reformulated within 
the context of the recently developed
formalism of geometrothermodynamics. This reformulation 
is shown to be invariant with respect to Legendre transformations,
and to allow several equivalent representations.  
Legendre invariance allows
us to explain a series of contradictory results known in the
literature from the use of Weinhold's and Ruppeiner's 
thermodynamic metrics for black holes. 
For the Reissner-Nordstr\"om black hole the geometry
of the space of equilibrium states  
is curved, showing a non trivial thermodynamic interaction,
and the curvature contains information about critical points 
and phase transitions. On the contrary, for the Kerr black hole the geometry is 
flat and does not explain its phase transition structure.

\end{abstract}

\maketitle

\section{Introduction}
\label{sec:int}
The geometry of thermodynamics has been the subject of moderate research
since the original works by Gibbs \cite{gibbs} and
Caratheodory \cite{car}. Results have been achieved in 
two different approaches.  
The first one consists in introducing metric structures on the space
of thermodynamic equilibrium  states ${\cal E}$, whereas the second 
group uses the contact structure of the so-called thermodynamic phase space 
${\cal T}$. 
Weinhold \cite{wei1} introduced {\it ad hoc} on ${\cal E}$ a metric 
defined as the Hessian  of the internal thermodynamic energy, where the derivatives
are taken with respect to the extensive thermodynamic variables. 
Ruppeiner
\cite{rup79}
introduced a metric which is conformally equivalent to Weinhold's metric,
with the inverse of the temperature as the conformal factor. 
Results of the application of Ruppeiner's geometry have been reviewed in
\cite{rup95,john03,jan04,san05c}. This approach has found
applications also in the context of thermodynamics of black holes 
\cite{caicho99,shen05,aman03,aman06a,aman06b,sarkar06}. 

The second approach, developed
specially by Hermann \cite{her} and Mrugala
\cite{mru1,mru2}, uses the natural contact structure of the phase space ${\cal T}$.
Extensive and intensive thermodynamic
variables are taken together with the thermodynamic potential to
constitute well-defined coordinates on ${\cal T}$.  
A subspace of ${\cal T}$ is the space of thermodynamic equilibrium states
${\cal E}$, defined by means of a smooth embedding mapping $\varphi: {\cal E}
\longrightarrow {\cal T}$. This implies
that each system possesses its own space ${\cal E}$. On the other hand, on ${\cal T}$
it is always possible to introduce the fundamental Gibbs 1-form which, when projected
on  ${\cal E}$ with the pullback of $\varphi$, generates the first law of thermodynamics
and the conditions for thermodynamic equilibrium.
Furthermore,  on ${\cal T}$ it is also possible to consider 
Riemannian structures \cite{tormon93,herlac98}.

Geometrothermodynamics (GTD) \cite{quezar03,quev07} was recently developed as 
a formalism that  
unifies the contact structure on ${\cal T}$ with the metric structure 
on ${\cal E}$ in a consistent manner, by considering only 
Legendre invariant metric structures on both ${\cal T}$ and ${\cal E}$. 
This last property is important in order to guarantee that the thermodynamic 
characteristics of a system do not depend on the thermodynamic potential 
used for its description. One simple metric has been used in GTD in order
to reproduce geometrically the non critical and critical behavior of the
ideal and van der Waals gas, respectively. 
 In the present work we present a further application of GTD in 
general relativity, namely, we reformulate black hole thermodynamics and 
try to reproduce the phase transition structure  
of black holes by using one of the simplest metric structures that are 
included in GTD.  

In general relativity, the gravitational field of the 
most general black hole is described by the 
Kerr-Newman \cite{solutions} solution that corresponds to a rotating, 
charged black hole. 
The discovery by Bekenstein \cite{bek73} that the behavior of the horizon area of a black 
hole resembles the behavior of the entropy of a classical thermodynamic system
initiated an intensive and still ongoing investigation of what is now called thermodynamics
of black holes \cite{bch73,haw75,davies}. 
Several attempts have been made in order to describe the thermodynamic behavior of black 
holes in terms of metrics defined on ${\cal E}$ 
\cite{shen05,aman03,aman06a,aman06b,sarkar06}. 
In particular, Weinhold's and Ruppeiner's metrics were used to find a direct relationship 
between curvature singularities 
and  divergencies of the heat capacity. Unfortunately, the 
results lead to completely contradictory statements. For instance, for
the Kerr black hole Weinhold's metric predicts no phase transitions at all \cite{aman03},
whereas Ruppeiner's  metric, with  a very specific thermodynamic potential,
predicts phase transitions which are compatible with the results of 
standard black hole thermodynamics \cite{shen05}. It is one of the goals 
of this work to explain this contradiction by using an invariant approach. 
We will conclude that the 
origin of this inconsistency is due to the fact that Weinhold's and
Ruppeiner's metrics are not Legendre invariant, a property that makes 
them inappropriate for describing the geometry of thermodynamic systems.
From the vast number of Legendre invariant metrics
which are allowed in the context of GTD we choose probably the 
simplest one. This choice allows us to 
find Legendre invariant generalizations of Weinhold's and Ruppeiner's metrics.
In the case of two-dimensional GTD, we apply these 
Legendre invariant metrics, and obtain consistent
results.

This paper is organized as follows. In Section \ref{sec:gtd} we briefly 
review the fundamentals of GTD, and present a simple Legendre invariant 
metric. In Section \ref{sec:bht} we apply GTD to thermodynamics of black 
holes in general, and find the simplest Legendre invariant generalizations
of Weinhold's and Ruppeiner's metrics. 
In Sections  \ref{sec:rn} and \ref{sec:kerr}  we analyze the geometry of the 
Reissner-Nordstr\"om and Kerr black hole thermodynamics, by using Legendre 
invariant thermodynamic metrics.  
We show that in both cases the results are geometrically consistent. 
We find an agreement with the results of standard black hole thermodynamics
in the case of the Reissner-Nordstr\"om solution. However, for Kerr black holes
we show that the simplest Legendre invariant metrics do not reproduce the 
corresponding phase transition structure. Section \ref{sec:firao} contains 
a brief analysis of the Fisher-Rao metric. 
 Finally, Section \ref{sec:con} is devoted
to discussions of our results and suggestions for further research.
Throughout this paper we use units in which $G=c=k_{_B}=\hbar =1$.

\section{Review of geometrothermodynamics}
\label{sec:gtd}
Consider the $(2n+1)$-dimensional thermodynamic phase space ${\cal T}$ 
coordinatized by the thermodynamic potential $\Phi$, extensive variables $E^a$, 
and intensive variables $I^a$ $(a=1,...,n)$. Consider on ${\cal T}$ a 
non-degenerate metric $G=G(Z^A)$, with $Z^A=\{\Phi, E^a, I^a\}$, 
and the Gibbs 1-form $     
\Theta = d\Phi - \delta_{ab} I^a d E^b$, with $ 
\delta_{ab}={\rm diag} (1,1,...,1)$. 
The set $({\cal T},\Theta,G)$ 
defines a contact Riemannian manifold \cite{her,herlac98} if the condition
$\Theta \wedge (d\Theta)^n \neq 0$ is satisfied. Moreover, the 
metric $G$ is Legendre invariant if its functional dependence on $Z^A$ 
does not change under a Legendre transformation \cite{arnold}. 
The Gibbs 1-form $\Theta$ is also invariant with respect to
Legendre transformations. Legendre invariance guarantees that the geometric properties 
of $G$ do not depend on the thermodynamic potential used in its construction. 

The $n$-dimensional subspace ${\cal E} \subset {\cal T}$ determined by the smooth mapping 
$ \varphi : \   {\mathcal E} \  \longrightarrow {\mathcal T}$, that in terms 
of coordinates reads $ \varphi :  (E^a) \longmapsto (\Phi, E^a, I^a)$ 
with $\Phi=\Phi(E^a)$, is called the space of equilibrium thermodynamic
states if the condition $\varphi^*(\Theta)=0$ is satisfied, i.e.
\be
d\Phi = \delta_{ab} I^a d E^b \ , \qquad \frac{\partial\Phi}{\partial E^a} = 
\delta_{ab} I^b \ .
\ee
The first of these equations corresponds to the first law of thermodynamics, whereas
the second one is usually known as the condition for thermodynamic equilibrium \cite{callen}.
In the GTD formalism, the last equation also means that 
the intensive thermodynamic variables are dual to the extensive ones. Notice that
the mapping $\varphi$ as defined above implies that the equation $\Phi=\Phi(E^a)$ 
must be explicitly given. In standard thermodynamics this is known as the 
fundamental equation from which all the equations of state can be derived 
\cite{her,callen}. In this representation, the second law of thermodynamics is equivalent
to the convexity condition on the thermodynamic potential 
$ \partial^2 \Phi/\partial E ^a \partial E ^b \geq 0$ \cite{burke,callen}.

The thermodynamic potential satisfies  the homogeneity condition 
$\Phi(\lambda E^a) = \lambda^\beta \Phi(E^a)$ for constant parameters $\lambda$ and
$\beta$. Using the first law of thermodynamics, it can easily be shown that this 
condition leads to the relations 
\be
\beta \Phi(E^a) = \delta_{ab}I^b E^a \ , \qquad
(1-\beta)\delta_{ab}I^a d E^b +\delta_{ab} E^a d I^b = 0 \ ,
\ee
which are known as Euler's identity and Gibbs-Duhem relation.

The final ingredient of GTD is a non-degenerate metric structure $g$ on ${\cal E}$ 
from which we demand to be compatible with the metric $G$ on ${\cal T}$. This can 
be reached by using the pullback $\varphi^*$ in such a way that $g$ becomes 
naturally induced by $G$  as $g= \varphi^*(G)$. 
As shown in \cite{quev07}, a Legendre invariant metric $G$ induces a Legendre invariant
metric $g$. Vice versa, a metric $g$ on ${\cal E}$ is Legendre invariant only if
it is induced by a Legendre invariant metric $G$ on ${\cal T}$. It is in this sense that
one can show that Weinhold's and Ruppeiner's metrics, which are defined on ${\cal E}$,
are not Legendre invariant. Nevertheless, there is a vast  number of metrics on ${\cal T}$
that satisfy the Legendre invariance condition. For instance, the metric structure
\be
G = \Theta^2 + (\delta_{ab}E^a I^b)(\delta_{cd}d E^c d I^d)  \ ,
\label{mont}
\ee
where $\Theta$ is the Gibbs 1-form,  is Legendre invariant and induces on ${\cal E}$
the metric 
\be
g=\Phi \frac{\partial^2\Phi }{\partial E^a \partial E^b} dE^a d E^b \ .
\label{simpleg}
\ee
An important feature of this metric is that it is flat for an ideal gas and
non-flat for the van der Waals gas, with curvature singularities at the critical 
thermodynamic points \cite{quev07}. This is an indication that it can be used 
as a Legendre invariant measure of thermodynamic interaction. Although this 
property is shared by other metrics on ${\cal E}$ in the following analysis we
will use the specific choice (\ref{simpleg}) because of its simplicity.

Finally, we mention that the geometry of the metric $g=\varphi^*(G)$ is invariant
with respect to arbitrary diffeomorphisms performed on ${\cal E}$. This can 
be shown by considering explicitly the components of $g$ in terms of the components of $G$,
and applying arbitrary Legendre transformations on $G$. This important property 
allows us to consider variational principles in GTD that impose additional 
conditions on the metric structures \cite{quevaz07}.

\section{Black hole thermodynamics}
\label{sec:bht}
Vacuum black holes in Einstein's theory 
are completely characterized by the mass $M$, angular momentum
$J$, and electric charge $Q$. Although the statistical origin 
is still obscure, black holes possess thermodynamic properties 
specified through Hawking's temperature $T$, proportional to
the surface gravity on the horizon, and entropy $S$ proportional to
the horizon area \cite{bek73,haw75}. All these parameters are related by means of 
the first law of black hole thermodynamics $  
dM = T dS + \Omega_H  d J + \phi d Q$ (see, for instance, \cite{bch73}), 
where $\Omega_H $ is the angular velocity on the horizon, and $\phi$ is the
electric potential. For a given fundamental equation $M=M(S, J, Q)$ 
we have the conditions for thermodynamic equilibrium
\be
T =\frac{\partial M}{\partial S} \ , \quad
\Omega_H  =\frac{\partial M}{\partial J} \ , \quad
\phi =\frac{\partial M}{\partial Q} \ .
\ee
Thus, the phase space ${\cal T}$
 for black hole thermodynamics is  7-dimensional with coordinates
$Z^A=\{M,S,J,Q,T,\Omega_H ,\phi\}$. 
The fundamental Gibbs 1-form is given by $\Theta = dM - T dS - \Omega_H  d J - \phi d Q$. 
The space of thermodynamic equilibrium states ${\cal E}$ is 3-dimensional 
with coordinates $E^a = \{S,J,Q\}$, and is defined by means of the mapping
\be
\varphi: \{S,J,Q\} \longmapsto \left\{ M(S,J,Q), S,J,Q, \frac{\partial M}{\partial S}, 
\frac{\partial M}{\partial J}, \frac{\partial M}{\partial Q}\right\} \ .
\ee
The mass $M$ plays the role of thermodynamic potential that depends on the
extensive variables $S$, $J$ and $Q$. However, Legendre transformations  
allow us to introduce a set of seven additional thermodynamic potentials which 
depend on different combinations of extensive and intensive variables. 
The complete set of thermodynamic potentials
can be written as
\bea
M & = & M(S,J,Q) \ , \nonumber \\
M_1 & = & M_1(T,J,Q) = M - TS \ ,\nonumber \\ 
M_2 & = & M_2(S,\Omega_H ,Q) = M - \Omega_H  J \ ,\nonumber \\ 
M_3 & = & M_3(S,J,\phi) = M - \phi Q \ ,\nonumber\\ 
M_4 & = & M_4(T,\Omega_H  ,Q) = M - TS -\Omega_H  J \ , \\ 
M_5 & = & M_5(T,J,\phi) = M - TS -\phi Q\ ,\nonumber\\ 
M_6 & = & M_6(S,\Omega_H  ,\phi) = M - \Omega_H  J -\phi Q \ ,\nonumber\\
M_7 & = & M_7(T,\Omega_H  ,\phi) = M - TS -\Omega_H  J -\phi Q \ .\nonumber 
\label{tpotentials}
\eea
Notice that the mapping $\varphi$ can be defined in each case, independently 
of the chosen thermodynamic potential. On the other hand, since we are considering
only Legendre invariant structures on ${\cal T}$ and ${\cal E}$, the characteristics
of the underlying geometry for a given thermodynamic system 
will be independent of the thermodynamic potential. This is in agreement 
with standard thermodynamics.   
Consequently, in the mass representation of black hole thermodynamics 
described above, we have the freedom of choosing anyone of the potentials 
$M, M_1, ..., M_7$, without affecting the thermodynamic properties of black
holes. 

In the context of GTD, it is also possible to consider the entropy representation.
In this case, the Gibbs 1-form of the phase space can be chosen as 
\be
\Theta_S =  dS -\frac{1}{T} dM  +\frac {\Omega_H }{T} d J + \frac{ \phi}{T} d Q \ .
\ee
The space of equilibrium states is then defined by the smooth mapping
\be
\varphi_S: \{M,J,Q\} \longmapsto \left\{ M, S(M,J,Q),J,Q, T(M,J,Q), \Omega_H (M,J,Q),
\phi(M,J,Q)\right\} \ ,
\ee
with
\be
\frac{1}{T} = \frac{\partial  S}{\partial M} \ ,\quad
\frac{\Omega_H }{T} = -\frac{\partial  S}{\partial J} \ ,\quad
\frac{\phi}{T} = -\frac{\partial  S}{\partial Q} \ , 
\ee
such that $\varphi^*_S(\Theta_S) =0$ leads to the first law. In the entropy representation
the fundamental equation is now given by $S=S(M,J,Q)$, and the second law of 
thermodynamics corresponds to the concavity condition of the entropy function. 
Additional representations
can easily be analyzed within GTD, and the only object that is needed in each case 
is the smooth mapping $\varphi$ which guarantees the existence of a well-defined 
space of equilibrium states. 
Clearly, the thermodynamic
properties of black holes must be independent of the representation. 

Now we consider metric structures on ${\cal E}$. For black holes, 
Weinhold's metric $g^W$  is defined as the Hessian in the mass representation \cite{wei1},
whereas Ruppeiner's metric $g^R$ is given as minus the Hessian in the entropy
representation \cite{rup79}. From the analysis given above, it is clear that these metrics
must be related by $g^W = T g^R$. As we showed in \cite{quev07},
the main problem with Weinhold's and Ruppeiner's metrics is that they are
not Legendre invariant. In GTD it is possible to derive, in principle, 
an infinite number of metrics which preserve Legendre invariance; nevertheless,
according to Eq.(\ref{simpleg}), the simplest way to reach the Legendre invariance 
for $g^W$ is to apply a conformal transformation, with the thermodynamic potential 
as the conformal factor. Consequently, the simplest Legendre invariant 
generalization of Weinhold's metric can be written in components as   
\be
g= M g^W = M \frac{\partial^2 M}{\partial E^a \partial E^b} d E^a d E ^b \ , 
\label{gbhe}
\ee
where $E^a=\{S,J,Q\}$. This Legendre invariant metric can also be 
written in terms of the components of Ruppeiner's metric as
\be
g = M T g^R = - M \left(\frac{\partial S}{\partial M}\right)^{-1}
 \frac{\partial^2 S}{\partial F^a \partial F^b} d F^a d F^b \ ,
\label{rup}
\ee
with $ F^a = \{M,J,Q\}$. Using the mass representation, 
in the phase space ${\cal T}$  the corresponding generating metric structure 
can be written as  [cf. Eq.(\ref{mont})]
\be
G = (dM - TdS - \Omega_H  d J -\phi d Q)^2 + (TS+\Omega_H  J + \phi Q) ( dT d S + d\Omega_H  d J + d\phi d Q) \ .
\label{gbht}
\ee
Notice that to obtain (\ref{gbhe}) we need to use Euler's identity for the
conformal factor in front of the second term in (\ref{gbht}), i.e. 
$\beta M =  TS+\Omega_H  J + \phi Q$. Thus, $g$ as given in (\ref{gbhe}) 
is determined only up to the multiplicative constant $\beta$ that, obviously,
does not affect its geometry.   

In the following sections we will analyze metrics (\ref{gbhe}) and (\ref{rup})  
in the case of 2-dimensional GTD 
with $a,b=1,2$, $E^1 =S$, and $E^2$ will be chosen either as $Q$ or as $J$,
which corresponds to the Reissner-Nordstr\"om and Kerr black holes, respectively.

\section{The Reissner-Nordstr\"om black hole}
\label{sec:rn}
The Reissner-Nordstr\"om solution describes a static black hole with mass $M$ 
and electric charge $Q$. 
The inner 
and outer event horizons are situated at $r_-$ and $r_+$ so that the outer horizon
area is $A=4\pi r_+^2$, where  $r_\pm = M\pm \sqrt{M^2-Q^2}$.  The extremal black hole corresponds to the value $r_+=r_-$
and we suppose that $M^2\geq Q^2$ in order to avoid naked singularities.
From the horizon area law it follows that the entropy of the black hole is given by
\be
S=\frac{1}{4} A = \pi \left(M+\sqrt{M^2-Q^2}\right)^2 \ ,
\label{srn}
\ee
an expression that can be rewritten as 
\be
M=\frac{1}{2 \sqrt{\pi S}}\left(\pi Q^2+ S \right) \ .
\ee
This is the fundamental equation from which, according to Eq.(\ref{gbhe}), 
one easily calculates the Legendre invariant 
metric on the space of equilibrium states. Then
\be
g= \frac{1}{S^2}\left( \pi Q^2 + S\right)\left[\frac{1}{16\pi S}   
\left(3\pi Q^2 - S \right) dS^2 -\frac{1}{2}Q dS dQ + \frac{1}{2} S dQ^2\right]
\ .
\label{metrn}
\ee
The convexity condition is not satisfied in general. Only the
term $g_{QQ}$ is always positive definite, whereas $g_{QS}$ is positive only 
for negative values of the total charge. The component $g_{SS}$ violates the
convexity condition if $S > 3\pi Q^2$. The limiting value $S = 3\pi Q^2$
determines the turning point where the second law of thermodynamics becomes invalid.
In terms of the horizons' radii  and for a given radius $r_+$ of the outer horizon, 
this is equivalent to the statement that the  convexity condition is 
valid only in the interval $r_- \in\  [r_+/3, r_+)$. 
 
The scalar curvature corresponding to the metric (\ref{metrn})  reads
\be
R = - \frac{8 \pi^2 Q^2 S^2 (\pi Q^2 -3S)}{(\pi Q^2 +S)^3 (\pi Q^2 -S)^2 } \ .
\ee
We see that the only curvature singularity occurs when $S=\pi Q^2$. 
This corresponds to 
the value $M=Q$, i. e. the extremal black hole. We interpret this result
as an indication of the limit of applicability of GTD as a geometric model
for equilibrium thermodynamics. This is also in accordance with the
intuitive expectation that naked singularities show the limit
of applicability of black hole thermodynamics. 

Another interesting point is $S =\pi Q^2/3$, where the scalar curvature 
vanishes identically, leading to a flat 
geometry. At this point the scalar curvature changes its sign, and it is the
only point where this happens. 
Notice that the value $S =\pi Q^2/3$ corresponds to 
$M= 2Q/\sqrt{3}$ or, equivalently, $r_+=3r_-$ which according to Davies \cite{davies}
is exactly the point where the system is undergoing a phase transition. 

It is interesting to note that the phase transition point has been analyzed 
in other works, using Weinhold's and Ruppeiner's metrics, with 
partially contradictory results. For instance, in \cite{aman03} at the phase 
transition point $S =\pi Q^2/3$ nothing happens because Ruppeiner's metric is flat 
everywhere. On the other hand, in \cite{shen05} this point corresponds to a true
curvature singularity of Ruppeiner's geometry with a different thermodynamic potential.
Moreover, in the same work the extremal black hole is described 
by a well-behaved metric with zero scalar curvature that, in principle,
can be analytically  extended to include the 
case of a naked singularity. We interpret these contradictory results as due to the
use of metrics that are not Legendre invariant.  In fact, let us consider the
simplest Legendre invariant generalization  of Ruppeiner's metric (\ref{rup})
with the fundamental equation (\ref{srn}). Then 
\bea
g= \frac{\pi M }{(M^2-Q^2)S }\bigg\{&& \left[ 2 M^3 -3MQ^2 + 2(M^2-Q^2)^{3/2}\right]
dM^2 \nonumber \\
&& +  2 Q^3 dM d Q - \left[M^3 + (M^2-Q^2)^{3/2} \right] dQ^2\bigg\} \ . 
\eea
From the corresponding scalar curvature one can see that it diverges at $M=Q$,
and changes its sign at $M=2Q/\sqrt{3}$,  which coincides with the result 
obtained by using the Legendre invariant Weinhold metric 
given in Eq.(\ref{metrn}). This solves the incompatibility problem between 
the results obtained by using geometries which do not preserve 
Legendre invariance. 

What we learn in this case from the use of  Legendre invariant metrics is that 
in GTD a phase transition can also be described by a change of sign of the 
scalar curvature, passing through a state of flat geometry. Although there
is no singular behavior associated with this phase transition, we believe that
the change of topology that occurs when going from a negative to a positive
curvature could have drastic consequences for the underlying thermodynamics.
A more detailed analysis will be necessary in order to clarify this issue.

\section{The Kerr black hole}
\label{sec:kerr}
The Kerr solution describes the gravitational field of rotating black hole 
with mass $M$ and angular momentum $J$. 
The inner and outer horizons are situated at $r_-$ and $r_+$,
where $r_\pm = M \pm \sqrt{M^2-a^2}$. 
 The entropy
is calculated as usual in terms of the area of the horizon
\be
S=\frac{1}{4} A =  2\pi \left( M^2 + \sqrt{M^4-J^2}\right) 
\label{feskerr}
\ee
which for the mass representation can be rewritten as
\be
M = \sqrt{\frac{S}{4\pi} + \frac{\pi J^2}{S} } \ .
\ee
From Eq.(\ref{gbhe}) we get the corresponding Legendre invariant 
metric of the space of thermodynamic equilibrium states
\be
g = \frac{\pi S}{ S^2 + 4\pi^2 J^2}\left[
\left( \frac{3\pi^2 J^4}{S^4} + \frac{3J^2}{2 S^2} - \frac{1}{16 \pi^2}\right) d S^2
- \frac{ J}{S^3}\left(3 S^2 + 4 \pi^2 J^2 \right) dJ d S + dJ^2\right] \ .
\label{gmkerr}
\ee
Unexpectedly, the curvature of this metric vanishes. The same result was obtained
in \cite{aman06a} by using Weinhold's metric. This is a surprising result because
it would mean that  Kerr black holes do not show any statistical thermodynamic 
interaction. On the other hand, the standard thermodynamics of Kerr black holes is by 
no means trivial and shows a very rich phase transition structure \cite{davies}.
Moreover, the analysis performed in \cite{shen05} using Ruppeiner's metric 
closely reproduces this structure. This contradictory result is again due 
to the use of non invariant metrics. In fact, if we consider in GTD the simplest
Legendre invariant generalization of Ruppeiner's metric as given in Eq.(\ref{rup}),
we obtain from the fundamental equation (\ref{feskerr}) the following metric
\be
g= \frac{2\pi}{S (M^4-J^2) }
\left\{\left[ (M^4-J^2)^{3/2}+M^2(M^4-3J^2)\right] dM^2 + 2M^3J dMdJ - 
\frac{M^4}{2}dJ^2 \right\} \ ,
\ee
whose curvature also vanishes. This coincides with the result obtained by using
the Legendre invariant generalization (\ref{gmkerr}) of Weinhold's metric, but
it also drastically differs from the result of \cite{shen05}
where a non zero curvature was obtained for the pure Ruppeiner metric with a special choice
of the thermodynamic potential; instead of using $M$, the authors of \cite{shen05} define an 
internal energy which in the notation used here corresponds to the entropy representation 
of the  thermodynamic potential $M_2$ given in Eq.(13). One can easily 
show that the use of the original potential $M$ leads to a different metric with non
vanishing curvature, and the only curvature singularity appears at $M^2=J$, i.e. in the
extremal black hole limit, a result that does not reproduce the phase transition 
structure of the Kerr black hole. This shows that Ruppeiner's metric leads to completely
different results, depending on the thermodynamic potential.  

The main result of this Section is that Weinhold's and Ruppeiner's metrics in their
Legendre invariant generalizations lead to the same result for the Kerr black hole.
Unfortunately, the corresponding geometry in the space of equilibrium states is flat
and does not reproduce the phase transition structure of the black hole. This is a
negative result which calls for a reconsideration of the use of geometric structures
in black hole thermodynamics.

\section{The Fisher-Rao metric}
\label{sec:firao}
In classical statistical mechanics, an alternative approach has been used 
to analyze the geometry of thermodynamic systems. The starting 
point is the probability density distribution which is given by the Gibbs measure
\be
p(x|\theta) = \exp\left[ - \theta_i H_i(x) - \ln Z(\theta)\right] \ ,
\label{prob}
\ee
where $Z(\theta)$ is the partition function, $H_i(x)$ are Hamiltonian 
functions and $\theta^i$, $i=1,...,n$ represent the $n$ 
parameters characterizing the 
statistical model under consideration. 
It can be shown \cite{brohug1} that for each value of the parameters 
the square root of this density can be associated to a vector in the 
Hilbert space ${\cal H}$. Consequently, ${\cal H}$ contains the state 
space of the system, and the properties of the statistical system  can 
be described by means of the embedding of $p(x|\theta)$ in ${\cal H}$.
Once the Hilbert space is considered in the language of projective 
geometry, it is possible to generalize this embedding construction 
to include the cases of  
quantum mechanical dynamics of equilibrium  states and pure quantum 
mechanics \cite{brohug2,brohug3}. 
The geometry that arises from the embedding turns out to possess
a natural Riemannian metric, the Fisher-Rao metric in the classical
case or the Fubini-Study metric in the quantum case. 

For the classical Gibbs distribution (\ref{prob}) the Fisher-Rao metric
takes the simple form \cite{brorit}
\be
g^{FR}= \frac{\partial^2 \ln Z(\theta)}{\partial \theta^i \partial \theta^j} d\theta^i 
d\theta^j \ .
\label{fisrao}
\ee
The geometric properties of the manifold described by this metric has been analyzed 
for different statistical models. In the case of the van der Waals gas the 
parameters can be chosen as $\theta^1 = 1/T$, and $\theta^2 = P/T$, and the 
corresponding Hamiltonian 
functions are the internal energy $U$ and volume $V$, respectively, so that 
$Z(\theta)$ is a function of temperature and pressure. The scalar curvature of 
this two-dimensional manifold turns out to diverge at the critical points, and 
the scaling exponent of the curvature near the transition points coincides with
that of the correlation volume \cite{rup95}. Furthermore, in the limiting 
case of an ideal gas, the curvature vanishes and the manifold is flat. 
This is exactly the behavior shown by Ruppeiner's and Weinhold's geometry in 
these particular cases. In fact, it can be shown \cite{jjk03}  
that in general both metrics are related to the Fisher-Rao metric by means of 
Legendre transformations of the corresponding variables. This explains why 
their geometric properties are similar, and indicates that the Fisher-Rao
metric is also not Legendre invariant.

To be more specific in the case under consideration in this work, we first 
mention that, due to its statistical origin, the components of the 
Fisher-Rao metric $g^{FR}_{ij}(\theta)$ are usually
given in terms of the ``inverse" of the thermodynamic variables: $\theta^1=1/T$,
$\theta^2=P/T$, etc. Since the relationships $\theta^i=\theta^i(E^a)$ must allow
the inverse transformation, it is easy to show that in the coordinates used here
the Fisher-Rao metric can be written as $g^{FR}_{ab} = \partial^2 \ln Z(E)/\partial E^a
\partial E^b$. The partition function for black holes is given by (see, for instance,
\cite{johnson}) 
\be
Z= \exp\left[ -\frac{1}{T}(M-TS - \Omega_H J - \phi Q)\right] \ .
\ee
It then follows that for the Reissner-Nordstr\"om black
hole $Z=\exp(-M_5/T)$ and for the Kerr black hole $Z=\exp(-M_4/T)$, where
the thermodynamic potentials $M_4$ and $M_5$ are related to the mass representation
we are using here by the Legendre transformations given in Section \ref{sec:bht}.
Consequently, the components of the Fisher-Rao metric for black holes are 
essentially given by
$g^{FR}_{ab}= -\partial^2 (M/T) /\partial E^a \partial E^b$. We now briefly explain how 
to show that this metric is not Legendre invariant. According to GTD, there must exist in the thermodynamic phase space ${\cal T}$ 
a metric $G^{FR}$ which generates $g^{FR}$ by means of the pullback $g^{FR}= \varphi^*(G^{FR})$.
On ${\cal T}$ we perform an arbitrary Legendre transformation 
$Z^A \rightarrow \tilde Z ^A$ which 
when acting on $G^{FR}$ produces the Legendre transformed metric  
$\tilde G ^{FR}$. Then the Legendre
transformed Fisher-Rao metric $\tilde g ^{FR}$
 in the space of equilibrium states ${\cal E}$ is computed by 
 $\tilde g ^{FR} = \tilde \varphi ^*(\tilde G ^{FR})$,\
where $\tilde \varphi$ is the embedding mapping in the new coordinates 
(for more details see \cite{quev07}). 
As a result we obtain that the functional dependence of $\tilde g ^{FR}$ is 
completely different from that of $g ^{FR}$; i.e. the Fisher-Rao metric 
is not Legendre invariant.
Similar results can be obtained by using the original potentials $M_4$ and $M_5$.

The result of this Section is that the Fisher-Rao metric for black holes is not Legendre invariant,
and therefore cannot be used to solve the problem of contradictory results following from the application
of Weinhold's and Ruppeiner's approaches.

\section{Conclusions}
\label{sec:con}

In this work we formulated the thermodynamics of general relativistic
black holes in the language of geometrothermodynamics. The general 
thermodynamic phase space ${\cal T}$ turns out to be 7-dimensional
and it is always possible to introduce a smooth mapping from the 
3-dimensional space of thermodynamic equilibrium states ${\cal E}$ 
to the phase space ${\cal T}$. Different formalisms
based on different representations, such as the energy or the entropy 
representation, can easily be handled within GTD. The equivalence 
between all possible representations is an obvious consequence of
the properties of the structures used in GTD. 
We present all the thermodynamic potentials that can be 
derived by means of arbitrary Legendre transformations, starting
from the mass representation, and explain why the geometric properties
of a thermodynamic system cannot depend on the chosen thermodynamic
potential. From all Legendre invariant metric structures that can
be introduced on ${\cal T}$ and, consequently, on ${\cal E}$,
we choose a simple example which allows
us to generalize other metrics used in the literature for investigating
the geometric properties of thermodynamic systems.

We studied two-dimensional GTD in the case of the Reissner-Nordstr\"om and
Kerr black holes, by using  simple Legendre invariant generalizations 
of Weinhold's and Ruppeiner's metrics. Our results show that for the 
thermodynamics of the Reissner-Nordstr\"om black hole there exists a
Legendre invariant geometry with non vanishing curvature. There is 
a true curvature singularity when the black hole becomes extremal. We 
interpret this result as indicating the limits of applicability of GTD
in the sense that the thermodynamic processes associated with 
the black hole becoming extremal must be highly non trivial and related 
to non equilibrium thermodynamics, an issue that has not yet
been considered within GTD. A second critical point occurs when $M=2Q/\sqrt{3}$
(i. e. $r_+ = 3 r_-$). At this point the scalar curvature changes its sign,
passing through a state of flat geometry. It also coincides with a thermodynamic
critical point where, according to Davies \cite{davies}, the system is
undergoing a phase transition. We propose that in GTD the  change of topology, 
that happens when the scalar curvature changes its sign, can be associated 
to a drastic change of the thermodynamic properties of the system, like 
a phase transition. This question needs to be further analyzed in order 
to get a more concrete answer. Our results also solve an incompatibility 
existing in the literature. Aman et al. \cite{aman03} used Ruppeiner's metric
for the thermodynamics of the Reissner-Nordstr\"om black hole 
to show that its geometry has no critical points which could be related to 
phase transitions, because it is a flat geometry. On the other hand, 
Shen et al. \cite{shen05} studied Ruppeiner's geometry and found a true
curvature singularity at $r_+ = 3r_-$, corresponding to a second order
phase transition. We proved that this contradiction is due to the 
use of metrics which do not preserve Legendre invariance. Our invariant
generalizations of Weinhold's and Ruppeiner's metrics lead to compatible
results and reinforce the prediction of a phase transition.

Our study of GTD in the case of the Kerr black hole also solves the 
contradiction in the results of Aman et al. \cite{aman03} and Shen et
al. \cite{shen05}. We proved that the Legendre invariant generalizations
of Weinhold's and Ruppeiner's metrics lead to the  result that the
underlying geometry is flat. This is a surprising result which  does not
coincide with the analysis  of standard black hole thermodynamics that
predicts the existence of phase transitions. A flat geometry implies that
there is no thermodynamic interaction and, consequently, no phase transitions
at all. This is a negative result which, in our opinion, implies 
that we should critically reconsider the application of geometry structures
in black hole thermodynamics. However, this negative
result can also be interpreted as implying 
that Weinhold's and Ruppeiner's metrics, even in 
their Legendre invariant version, are not suitable for describing the
thermodynamics of black holes. The statistical origin of the Fisher-Rao metric
could be thought to be an advantage, when compared with metrics introduced
{\it ad hoc}. However, we have seen that Legendre invariance is not a property 
of this statistical metric.  

An intriguing result was obtained by Shen et al. \cite{shen05} for the
Kerr black hole. They work in  the entropy representation 
of Ruppeiner's metric with a different thermodynamic 
potential. Instead of using the mass $M$ as an extensive variable, they consider 
the potential $M_2 = M - \Omega_H J$, and reproduce exactly the phase transition 
structure of the Kerr black hole \cite{davies}. Although we have seen that this result can 
drastically be changed by using $M$ as thermodynamic potential, it would be
interesting to consider the metric used by Shen et al. as a guide to find 
a generalization that would preserve Legendre invariance. 
 
In GTD there exists, in principle,
an infinite number of Legendre invariant metrics, and there is no reason 
to believe that all of them should be applicable to any thermodynamic system.
We think that it is necessary to find additional criteria which would serve
to select Legendre invariant metrics with certain specific properties. 
In this context, the application of variational principles in GTD could be 
useful. In fact, the metric induced on ${\cal E}$ by means of $g=\varphi^*(G)$
can be written in components as $(a,b=1,...,n,\ A,B =0,...,2n)$ 
\be
g_{ab} = \frac{\partial Z^A}{\partial E^a} \frac{\partial Z^B}{\partial E^b}
G_{AB} \ .
\ee
Then, we can limit ourselves, for example, to only those metrics $g_{ab}$
which define an extremal $n$-dimensional hypersurface 
on ${\cal E}$, i.e. metrics satisfying the ``motion equations" following
from the variation $\delta \int \sqrt{{\rm det}\,( g_{ab})} d^n x = 0$. 
This is equivalent to demanding that the mapping $\varphi:{\cal E}\longrightarrow
{\cal T}$ determines a non linear sigma model. The resulting geodesic-like 
equations can be solved for a given fundamental equation and we obtain as a result
the set of Legendre invariant 
metrics that can be used to describe the corresponding thermodynamic
system. Also, for a given Legendre invariant metric one can find the set 
of fundamental equations that satisfy the corresponding equations. 
This task is currently under investigation \cite{quevaz07}.

\section*{Acknowledgements} 
It is a great pleasure to dedicate this work to Bahram Mashhoon, on the
occasion of his 60-th birthday. 
I would like to thank F. Nettel and A. V\'azquez for 
helpful discussions and comments.
This work was supported in part by Conacyt, M\'exico, grant 48601.





\end{document}